\documentclass{article}
\usepackage{fullpage}	% I like previewing in this mode.  uncomment if you'd like. -DS
\usepackage{amsmath}
\usepackage{amssymb}
\usepackage{amsthm}
\usepackage{amssymb}  % assumes amsmath package installed
\newtheorem{thm}{Theorem}[section]
\newtheorem{lem}[thm]{Lemma}
\theoremstyle{remark}

\newcommand{\set}[1]{\left\{#1\right\}}
\newcommand{\segment}[1]{\ensuremath{\overline{#1}}}
\author{Deniz Sar{\i{}}{\"{o}}z\\City University of New York, The Graduate Center\\New York, NY 10016}
\title{Generalized Delaunay Graphs with respect to \\any Convex Set are Plane Graphs}
\begin{document}
%\keywords{Generalized Delaunay graph, Homothets, Scaled Translates, Geometric Graphs}
\maketitle
\thispagestyle{empty}
\begin{abstract}
We consider two types of geometric graphs on point sets on the plane based on a plane set $C$:
one obtained by translates of $C$, another by positively scaled translates (homothets) of $C$.
For compact and convex $C$, graphs defined by scaled translates of $C$, 
i.e., Delaunay graphs based on $C$, are known to be plane graphs.
We show that as long as $C$ is convex, both types of graphs are plane graphs.
\end{abstract}
\section{Introduction and Preliminaries}
%We consider two types of geometric graphs on point sets on the plane based on a convex set $C$:
%one due to intersecting translates of $C$, another due to intersecting positively scaled translates of $C$.
%The latter type includes the family of generalized Delaunay graphs for convex distance functions as a special case,
%namely, when $C$ is compact convex.
%In the case of not only convex but also compact $C$, the graph obtained by scaled translates of $C$ 
%has been referred to as the Delaunay graph based on $C$.
Let $P$ be a finite set of points on the Euclidean plane, and let $C$ be an arbitrary subset of the Euclidean plane.
Denote by $G_{t}(P, C)$ 
the graph on $P$ with edge set
consisting of those pairs of points
that are the intersection of $P$ with some translate of $C$.
Similarly, denote by $G_{st}(P, C)$
the graph on $P$ with edge set
consisting of those pairs of points
that are the intersection of $P$ with some positively scaled translate (homothet) of $C$.
We do not distinguish between graphs and their straight-line drawings.

If $C$ is a disk, notice that $G_{st}(P,C)$ is the usual Delaunay graph on $P$.
If $C$ is compact and convex, then $G_{st} (P,C)$ is the generalized Delaunay graph on $P$ with respect to (any convex distance function based on) $C$.
It is a triangulation if and only if no four points in $P$ lie on the boundary of a scaled translate of $C$.
If $C$ is a non-convex or non-compact set, then it can no longer be used for defining a distance function,
and in turn, it cannot be used for defining a Voronoi diagram.
Nonetheless, 
we still refer to $G_{st}(P,C)$ as the generalized Delaunay graph on $P$ with respect to $C$.

Recall that a graph drawn on the plane is called a plane graph if it satisfies the following conditions:
\begin{enumerate}
\item
no vertex is on an edge that it is not an endpoint of, and
\item
no edges cross, i.e., edges may only intersect at common endpoints.
\end{enumerate}

It was shown by Bose et al. \cite{Bose+10} that $G_{st}(P,C)$ is a plane graph in the case of compact and convex $C$.
Matou{\v s}ek et al. \cite{MSW90} have established planarity in an even more special case.
For any ordered pair $(P,C)$, notice that $G_{st}(P,C)$ has all the vertices and edges in $G_{t}(P,C)$,
so the latter must be a plane graph if the former is.
Our eventual goal is to show that the graphs $G_{st}$ are plane graphs whenever $C$ is convex, and not necessarily closed or bounded.
We will prove this first for $G_{t}$ graphs.

It is easy to argue that the plane graph condition 1 holds for $G_t$ and $G_{st}$ graphs based on any convex set $C$.
Since $C$ is convex, then every (possibly scaled) translate $C'$ of it is also convex.
Denote the line segment between two points $p$ and $q$ and inclusive of them as $\segment{pq}$.
For any three points $u, v, w \in P$ with $v \in \segment{uw}$, 
the convexity of $C'$ implies that it cannot meet $u$ and $w$ without meeting $v$,
so $C'$ will not induce the edge $uw$.
\mbox{Condition 2} remains to be shown for both the $G_{t}$ and the $G_{st}$ graphs.

\section{Geometric Graphs Defined by Translates of a Convex Set}
\begin{thm}\label{GtPlanar}
$G_{t}(P,C)$ is a plane graph for every finite point set $P$ whenever $C$ is convex.
\begin{proof}
Let $C$ be a convex subset of the plane.
Assume for contradiction that for some finite point set $P$, 
the graph $G_{t}(P,C)$ has an edge $ab$ that crosses another edge $cd$.
Then there are translates $C_1$ and $C_2$ of $C$ such that
$a,b \in C_1 \backslash C_2$ and $c,d \in C_2 \backslash C_1$.
Let $\mathbf{t}$ be the vector that $C_1$ can be translated by to obtain $C_2$.
Define the $+x$-axis in the direction of $\mathbf{t}$.

For the scope of this proof, for points
$p_1  (x_1, y)$ and 
$p_2  (x_2, y)$,
we write
$p_1 < p_2$
and $p_2 > p_1$
if $x_1 < x_2$.
Again for the scope of this proof,
for a point $p$ and a segment $\segment{qr}$ that it is outside of, 
if some point $s \in \segment{qr}$ shares the $y$-coordinate of $p$,
we write
$p < \segment{qr}$ if $p < s$, and
$p > \segment{qr}$ if $p > s$.

\begin{lem}
\label{SetDifferenceLeftRight}
If a point $p \in C_1 \backslash C_2$ and a point $q \in C_2$ have the same $y$-coordinate,
then $p < q$.
Similarly,
if a point $p \in C_2 \backslash C_1$ and a point $q \in C_1$ have the same $y$-coordinate,
then $p > q$.
\begin{proof}
Given a point $p \in C_1 \backslash C_2$ and a point $q \in C_2$ with the same $y$-coordinate,
assume for contradiction that $p < q$ does not hold.
Since $p$ and $q$ must be distinct, $p > q$.
But then we have a point 
$r = p+\mathbf{t} \in C_2$ by the definition of $\mathbf{t}$, we have $p < r$.
Now $p \in \segment{qr}$, so by convexity, $p \in C_2$,
a contradiction.
The second fact can be shown by a symmetric argument.
\end{proof}
\end{lem}

\begin{lem}
\label{noPointLeftBehind}
For every point $r \in C_1 \backslash C_2$ and points $p, q \subseteq C_2$, 
$r \not< \segment{pq}$.
(For every point $r \in C_2 \backslash C_1$ and points $p, q \subseteq C_1$, 
$r \not> \segment{pq}$.)
\begin{proof}
Given a point $r \in C_1 \backslash C_2$ and
a segment $\segment{pq} \subseteq C_2$, assume for contradiction that $r < \segment{pq}$.
Then there is a point $s \in \segment{pq}$ such that $r < s$.
By convexity, $s \in C_2$.
By the contrapositive of \ref{SetDifferenceLeftRight} we have 
$r \notin C_{1} \backslash C_{2}$.
The second fact can be proved by a symmetric argument.
\end{proof}
\end{lem}
To finish the proof of \ref{GtPlanar},
recall that $ab$ crosses $cd$ such that
$a,b \in C_{1} \backslash C_{2}$ and $c,d \in C_{2} \backslash C_{1}$.
Let $Q = \set{\ell, h}$ 
where $\ell$ and $h$ are
chosen from $\set{a,b,c,d}$ to respectively have minimum and maximum $y$-coordinate.
Denote $\set{a,b,c,d} \backslash Q$ by $\overline{Q}$.
This ensures that both points in
$\overline Q$ are comparable to $\segment{\ell h}$.
There are two possibilities.

Case 1: $Q=\set{a,b}$ or $Q=\set{c,d}$.
Let $\overline{Q} = \set{p,q}$.
The edge $pq$ crosses the edge $\ell h$, so without loss of generality, $p < \segment{\ell h} < q$.
Either $p$ or $q$ is in violation of \ref{noPointLeftBehind}, a contradiction.

Case 2: Without loss of generality, $Q = \set{a,c}$.
Still without loss of generality, $a=\ell$ and $c=h$.
Then since $ab$ and $cd$ cross, $b$ and $d$ have the same relation to $\segment{ac}$.
If $b < \segment{ac}$ and $d < \segment{ac}$, then since $ab$ crosses $cd$, we also have that $b < \segment{cd}$, 
which contradicts \ref{noPointLeftBehind}.
If $b > \segment{ac}$ and $d > \segment{ac}$, then since $ab$ crosses $cd$, we also have $b > \segment{cd}$, 
once again contradicting \ref{noPointLeftBehind}.
\end{proof}
\end{thm}

\section{Generalized Delaunay Graphs with respect to a Convex Set}
\begin{thm}\label{GstPlanar}
$G_{st}(P,C)$ is a plane graph for every finite point set $P$ whenever $C$ is convex.
\begin{proof}
Let $C$ be a convex subset of the plane.
Assume for contradiction that for some finite point set $P$, 
the graph $G_{st}(P,C)$ has an edge $ab$ that crosses another edge $cd$.
Then there are (possibly scaled) translates $C_1$ and $C_2$ of $C$ such that
$a,b \in C_1 \backslash C_2$ and $c,d \in C_2 \backslash C_1$.
If $C_2$ is a translate of $C_1$, we handled this case in the previous section.
Otherwise, without loss of generality, 
$C_2$ can be obtained by scaling $C_1$ by a factor $\alpha > 1$ about a point which we shall call $O$.

We now
set up a polar coordinate system centered at $O$ 
in which without loss of generality
\emph{every} point of $C_1$ is expressible as $(r,\theta)$ where $r>0$ and $\theta \in [-\pi/2, \pi/2)$.
Notice that $O \notin C_1$, otherwise we would have $C_1 \subseteq C_2$ due to convexity,
but we have $a, b \in C_1 \backslash C_2$ by hypothesis.
Since $O \notin C_1$, 
by the Separation Theorem for convex sets \cite{MatousekLDG},
some line $L$ through $O$ defines an open half-plane
that does not meet $C_1$.
Orient the plane such that $L$ is vertical with no point of $C_1$ to its left.
If $L$ meets $C_1$, then since $O \notin C_1$, by convexity,
$L \cap C_1$ must be above $O$, or below $O$.
Hence, without loss of generality, every point in $C_1$ can be expressed as  $(r,\theta)$ where $r>0$ and $\theta \in [-\pi/2, \pi/2)$.

Every point of $C_2$ is some point of $C_1$ scaled positively about $O$, 
so
it can be written as
$(r', \theta)$
where 
$\theta \in [-\pi / 2, \pi / 2)$
and $r' > 0$.
Moreover, every 
convex combination of two points from $C_1 \cup C_2$ has such a representation.
The following definitions, and the rest of the proof, are scoped to such points.

For the scope of \emph{this} proof, for points
$p_1 (r_1, \theta)$ and 
$p_2 (r_2, \theta)$,
we write
$p_1 < p_2$
and $p_2 > p_1$
if $r_1 < r_2$.
Again for the scope of \emph{this} proof,
for a point $p$ and a segment $\segment{qr}$ that it is outside of, 
if some point $s \in \segment{qr}$ shares the $\theta$-coordinate of $p$,
we write
$p < \segment{qr}$ if $p < s$, and
$p > \segment{qr}$ if $p > s$.

Now we prove the analogous fact to \ref{SetDifferenceLeftRight} whose proof is almost identical to \emph{its} proof.

\begin{lem}
\label{SetDifferenceLeftRightPolar}
If a point $p \in C_1 \backslash C_2$ and a point $q \in C_2$ have the same $\theta$-coordinate,
then $p < q$.
Similarly,
if a point $p \in C_2 \backslash C_1$ and a point $q \in C_1$ have the same $\theta$-coordinate,
then $p > q$.
\begin{proof}
Given a point $p \in C_1 \backslash C_2$ and a point $q \in C_2$ with the same $\theta$-coordinate,
assume for contradiction that $p < q$ does not hold.
Since $p$ and $q$ must be distinct, $p > q$.
But then we have a point 
$r = \alpha p \in C_2$, which means that $p < r$.
Now $p \in \segment{qr}$, so by convexity, $p \in C_2$,
a contradiction.
The second fact can be shown by a symmetric argument.
\end{proof}
\end{lem}

%For a point $p = (r, \theta)$ outside a segment $\segment{qs}$, 
%if there is a point $(r', \theta) \in \segment{qs}$,
%we write
%$p < \segment{qs}$ if $r < r'$, and
%$p > \segment{qs}$ if $r > r'$.
With these analogous re-definitions of $<$, the rest of the proof of \ref{GtPlanar} 
(including \ref{noPointLeftBehind} and its proof) can be repeated here, 
with the caveat that ``$\theta$-coordinate'' should be substituted in the place of ``$y$-coordinate''.
\end{proof}
\end{thm}

\section*{Acknowledgments}
J{\'a}nos Pach raised the question regarding the planarity of graphs defined by translates of convex shapes, 
probably as an exercise.
This work was supported by at least one of his grants.
Thanks also in order to Matias Korman for
informing me about related publications \cite{MSW90, Bose+10}
and encouraging me to publicize this.

\bibliographystyle{amsplain}
\bibliography{GenDel}

\end{document}